 \documentclass[preprint2]{aastex}
\usepackage[latin1]{inputenc}
\usepackage[T1]{fontenc}
\usepackage[english]{babel}
\usepackage{appendix}
\usepackage{color}

\usepackage{subfigure}
\usepackage{wrapfig}
\usepackage[babel]{csquotes}
\usepackage{natbib}
\usepackage{wasysym}
\usepackage{amssymb}
\usepackage{tabularx}
\usepackage{aas_macros}
\usepackage{color}

\shorttitle{{\it RedGOLD} Galaxy Cluster Detection in the NGVS}
\shortauthors{Licitra et al. 2016}

\begin{document}


\title{The Next Generation Virgo Cluster Survey. XX. {\it RedGOLD} Background Galaxy Cluster Detections}


\author{
Rossella Licitra\altaffilmark{1,2},
Simona Mei\altaffilmark{1,2,3},
Anand Raichoor\altaffilmark{1,4},
Thomas Erben\altaffilmark{5},
Hendrik Hildebrandt\altaffilmark{5},
Roberto P.~Mu\~noz\altaffilmark{6},
Ludovic~Van Waerbeke\altaffilmark{7},
Patrick~C\^{o}t\'{e}\altaffilmark{8},
Jean-Charles~Cuillandre\altaffilmark{9,1},
Pierre-Alain~Duc\altaffilmark{9},
Laura~Ferrarese\altaffilmark{8},
Stephen D.J.~Gwyn\altaffilmark{8},
Marc~Huertas-Company\altaffilmark{1,2},
Ariane~Lan\c{c}on\altaffilmark{10},
Carolina~Parroni\altaffilmark{1,2},
Thomas H.~Puzia\altaffilmark{6}
}

\altaffiltext{1}{GEPI, Observatoire de Paris, PSL Research University,  CNRS, University of Paris 
Diderot, 61, Avenue de l'Observatoire 75014, Paris France}
\altaffiltext{2}{University of Paris Denis Diderot, University of Paris Sorbonne Cit\'e (PSC), 75205 Paris Cedex
  13, France}
  \altaffiltext{3}{California Institute of Technology, Pasadena, CA 91125, USA}
  \altaffiltext{4}{CEA, Centre de Saclay, IRFU/SPP, 91191 Gif-sur-Yvette, France}
  \altaffiltext{5}{Argelander-Institut f\"{u}r Astronomie, University of 
Bonn, Auf dem H\"{u}gel 71, D-53121 Bonn, Germany}
\altaffiltext{6}{Institute of Astrophysics, Pontificia Universidad 
Cat\'{o}lica de Chile, Av. Vicu\~{n}a Mackenna 4860, 7820436 Macul, 
Santiago, Chile}
\altaffiltext{7}{Department of Physics and Astronomy, University of 
British Columbia, 6224 Agricultural Road, Vancouver, B.C., V6T 1Z1, Canada}
\altaffiltext{8}{Herzberg Institute of Astrophysics, National Research 
Council of Canada, Victoria, BC, V9E 2E7, Canada}
\altaffiltext{9}{CEA/IRFU/SAp, Laboratoire AIM Paris-Saclay, CNRS/INSU, Universit\'e Paris Diderot, PSL Research University, F-91191 Gif-sur-Yvette Cedex, France}
\altaffiltext{10}{Observatoire astronomique de Strasbourg, 
Universit\'{e} de Strasbourg, CNRS, UMR 7550, 11 rue de 
l'Universit\'{e}, F-67000 Strasbourg, France}


\begin{abstract}
We build a background cluster candidate catalog from the Next Generation Virgo Cluster Survey, using our detection algorithm RedGOLD. The NGVS covers 104~$deg^2$ of the Virgo cluster  in the $u^*,g,r,i,z$--bandpasses to a depth of $ g \sim 25.7$~mag (5$\sigma$). Part of the survey was not covered or has shallow observations in the $r$--band. We build two cluster catalogs: one using all bandpasses, for the fields with deep $r$--band observations ($\sim 20 \ deg^2$), and the other using four bandpasses ($u^*,g,i,z$) for the entire NGVS area. Based on our previous CFHT-LS W1 studies, we estimate that both of our catalogs are  $\sim100\%$($\sim70\%$) complete and $\sim80\%$ pure, at $z\le 0.6$($z\lesssim1$), for galaxy clusters with masses of $M\gtrsim10^{14}\ M_{\odot}$.   We show that when using four bandpasses,  though the photometric redshift accuracy is lower, RedGOLD detects massive galaxy clusters up to $z\sim 1$ with completeness and purity similar to the five--band case. This is achieved when taking into account the bias in the richness estimation, which is $\sim40\%$ lower at $0.5\le z<0.6$ and  $\sim20\%$ higher at $0.6<z< 0.8$, with respect to the five--band case. RedGOLD recovers all the X--ray clusters in the area with mass $M_{500} > 1.4 \times 10^{14} \rm M_{\odot}$ and $0.08<z<0.5$. Because of our different cluster richness limits and the NGVS depth, our catalogs reach to lower masses than the published redMaPPer cluster catalog over the area, and we recover $\sim 90-100\%$ of its detections. \end{abstract}





\section{Introduction}
Being galaxy clusters the largest gravitationally bound structures in the Universe, they represent a unique laboratory 
to study galaxy evolution and to quantify the importance of environmental effects on the evolution of galaxies.

Large area surveys are necessary to build a complete cluster catalog, useful for both constraints on cosmological parameters and the study of the cluster galaxy evolution. Since the available data sets cover large areas, automated cluster detection methods are required.

In this work, we present the catalog of background cluster candidates in the {\itshape Next Generation Virgo Cluster Survey} \citep[NGVS; ][]{Ferrarese2012}. The NGVS is a large program at the
{\itshape Canada France Hawaii Telescope} (CFHT), centered on M87 and imaging the Virgo cluster from
the inner regions  to its virial radius in five optical bands $u^*, g, r, i, z$. The NGVS covers 104 $\rm deg^2$ with a limiting magnitude $g \sim25.7$~mag \citep[5$\sigma$, 2" aperture; ][]{Raichoor2014}.

The main goal of the NGVS is to study the Virgo cluster at a depth never reached before in the optical   \citep[e.g.,][]{Ferrarese2012, Paudel2013, Raichoor2014}. The survey provides deep observations that are also useful for background science, such as galaxy and galaxy cluster studies, weak and strong lensing. 
Recently, we have published the homogenized  photometry and photometric redshift catalog of the background sources \citep{Raichoor2014}, which we use to obtain the results presented in this paper.

For the NGVS background cluster detections, we use the algorithm {\itshape RedGOLD}, presented in \citet{Licitra2016}, hereafter L16. 
Because of the need of automated cluster detection algorithms, a great effort has been made to develop codes based on different techniques and produce cluster catalogs, ideally including all the structures above a given mass threshold, with the lowest false detection rate. The best approach to detect different families of galaxy clusters is to combine catalogs built using different detection methods: for example, the cluster catalogs built using the Friends-of-Friends \citep[e.g.,][]{Wen2012} algorithm or the Voronoi tessellation \citep[][]{Ebeling1993} have the advantage of detecting structures with an irregular geometry. The red--sequence based cluster catalogs \citep[e.g, ][L16]{Gladders2000,Thanjavur2009, Rykoff2014} include relaxed and massive structures, while the matched filter cluster detection technique can be used to model peculiar characteristic of the galaxy clusters, such as the luminosity function or the cluster galaxy distribution \citep[e.g, ][L16]{Olsen2007, Grove2009, Milkeraitis2010, Bellagamba2011}. Each method is affected by false detections that contaminate their cluster catalogs, which can be minimized for each filter combination and survey characteristic. As discussed in L16, {\it RedGOLD} is a new cluster detection algorithm, based on a revised red--sequence technique.  It provides a richness estimate that tightly correlates with the cluster physical properties (e.g., temperature, mass).

The NGVS area has already been inspected by the cluster detection algorithm redMaPPer \citep{Rykoff2014}, using shallower optical data from the Sloan Digital Sky Survey \citep[SDSS;][]{York2000}.  redMaPPer searches for red galaxy overdensities and assigns a cluster redshift, richness, and likelihood.  In the SDSS, it successfully detects clusters with masses of  $M>10^{14}\ \rm M_{\odot}$ up to $z\sim 0.55$. In this paper, we will present the comparison of our cluster catalog with the redMaPPer detections up to intermediate redshifts.

The NGVS is only partially covered by deep $r$--band observations and we study the impact of the lack of the $r$--band on the {\it RedGOLD} richness estimation and its performance in terms of completeness and purity.  
This analysis will be useful also for future surveys with a inhomogeneous band coverage, such as the spatial Euclid mission \citep{Laureijs2011}. Euclid will cover an area of $15,000\ \rm deg^2$ in one broad optical band and in the near-infrared, and will need optical observations from ground--based surveys, such as the CFHT-LS \citep{Gwyn2012}, the {\it Large Synoptic Survey Telescope} \citep[LSST; ][]{lsstsb2012}, the {\it Dark Energy Survey} \citep[DES; ][]{DEScollab2005}, the {\it Panoramic Survey Telescope and Rapid Response System} \citep[Pan-STARRS; ][]{Kaiser2002}. As a consequence, different fields will have
a different band coverage and any bias due to these heterogeneous observations should be carefully understood.

This paper is organized as follows. In section~\ref{sec:Observations} we describe the observational data 
and the survey properties. We briefly present the photometric redshift estimation in section \ref{sec:photoz}, and in section \ref{sec:technique} we summarize our detection technique, extensively described in L16.
We present the cluster catalog obtained applying {\itshape RedGOLD} to the NGVS optical data in section \ref{sec:NGVSdetections}. 

We assume a standard cosmological model with $\Omega_m=0.3, \Omega_{\Lambda}=0.7$ and 
$H_0=70\ \rm km\ s^{-1}\ Mpc^{-1}$. The observed magnitudes are given in the AB system \citep{Oke1983,Sirianni2005} while the absolute rest-frame magnitudes are given in the Vega system.  We use the version dr8/v5.10 of the redMaPPer catalog.

\section{Observations and data description} \label{sec:Observations}

The NGVS observations were performed with the MegaCam instrument \citep{Boulade2003}, the optical imager mounted on MegaPrime, at 
the prime focus of the CFHT. In our analysis, we use the data reduction and photometric catalog from  \citet{Raichoor2014}. We refer to this work for further details. 

Briefly, the NGVS images were processed with the $ELIXIR$\footnote{http://www.cfht.hawaii.edu/Instruments/Elixir/} pipeline at the Canadian Astronomical Data Centre (CADC\footnote{http://www4.cadc-ccda.hia-iha.nrc-cnrc.gc.ca/cadc/}). The NGVS observations were obtained from 01/03/2008 until 12/06/2013,  under several CFHT programs (P.I. L. Ferrarese: 08AC16, 09AP03, 09AP04, 09BP03, 09BP04, 10AP03, 10BP03, 11AP03, 11BP03, 12AP03, 12BP03, 13AC02, 13AP03; P.I. S. Mei: 08AF20; P.I. J.-C. Cuillandre: 10AD99, 12AD99 and P.I. Y.-T. Chen: 10AT06). We reject images with an exposure time shorter than 100s, and obtained with unfavorable conditions at the CFHT. The astrometric and photometric calibration, and the image co-addition and mask creation are described in \citet{Raichoor2014} and follow the reduction procedures  adopted for the Canada-France-Hawaii Telescope Lensing Survey \citep[CFHTLenS; ][]{Erben2013}. Bright, saturated stars, and areas which would bias the analysis of faint background sources were masked as described in Raichoor et al. (2014). For the NGVS, a reliable masking of bright Virgo members is particularly important to obtain a homogeneous photometric catalog.

With respect to the standard THELI pipeline, our reduction was improved to obtain a homogeneous photometric calibration tied to the SDSS and realistic photometric  error estimates. The photometric catalogs  were obtained with the method described in \citet{Hildebrandt2012}, adopting the global point--spread--function (PSF) homogenization. This technique increases the quality of the photometric redshifts, because of the more accurate color estimations. 
Multi--wavelength catalogs were derived using SExtractor \citep{Bertin1996} in dual-image mode on each single pointing on the convolved images. The un--convolved $i$--band observations, having the better average seeing ($0.52\arcsec\pm0.04\arcsec$), have been chosen as detection images and the corresponding MAG\_AUTO provided by SExtractor has been adopted as the total $i$--band magnitude.

The total magnitudes in the $u^*grz$ bands are measured from the isophotal magnitudes, the SExtractor $MAG\_ISO$, as in \citet{Hildebrandt2012}:
 \begin{equation}
 {MAG_{j}}_{tot}=i_{tot}+(MAG\_ISO_{j}-MAG\_ISO_i)\ ,
 \end{equation}
 where $j=u^*, g, r, z$.
 
Photometric errors have been measured in the un--convolved images as described in \citet{Raichoor2014}, from the noise estimation in 2,000 random apertures in each bandpass and in each MegaCam pointing. In the un--convolved images, this corresponds to $\sim 1.5$ the photometric errors given by SExtractor \citep[see][for further details]{Raichoor2014}. A zero point uncertainty, estimated  comparing our photometry field--to--field to the SDSS, has been added in quadrature \citep[see also][]{Gwyn2012}.

The exposure time, depth and seeing for each bandpass are shown in Table \ref{tab:NGVSdepth}. The limiting magnitude is assumed as the $5\sigma$ detection limit in a 2'' aperture. 
The filter set is similar but not identical to the SDSS, and the conversion between the SDSS and MegaCam magnitudes is given by Eq.~4 in \citet{Ferrarese2012}.

The original NGVS observing strategy was to cover the whole area with the five bandpasses \cite[see ][]{Ferrarese2012}.
However, due to the exceptionally bad weather and dome shutter problems, the observations in the $r$--band are  available only in 34 out to 
117 MegaCam fields, which roughly correspond to $30\ \rm deg^2$, with eleven MegaCam fields shallower than originally planned in the $r$--band  (see \citealt{Ferrarese2012} and Fig.~1 in \citealt{Raichoor2014}).  For this reason, in Table \ref{tab:NGVSdepth}, for the $r$--band,  we provide the entire range of magnitude limits spanned in the 34 NGVS fields with the $r$-band data.

To calibrate the NGVS galaxy cluster detections, we use observations from the {\it Canada-France-Hawaii Telescope
Legacy Survey}  \citep[CFHT-LS;][] {Gwyn2012} Wide~1 (W1) field, and a reprocessed CFHTLenS reduction \citep{Erben2013}, described in \citet{Raichoor2014} and L16. The CFHT-LS Wide covers $154\ {\rm deg^2}$ in 5 optical bands, $u^*, g, r, i, z$, observed with the MegaCam instrument \citep{Boulade2003}, at a depth of  $i\sim25.7$~mag.
 The CFHTLenS photometry was obtained with the THELI pipeline \citep{Erben2013}, and photometric redshift measurements with PSF-matched photometry \citep{Hildebrandt2012}, such as for the NGVS reduction described above. 
We analyzed 62 out of the 72 pointings of the CFHT-LS W1, because we calibrated the photometric zero points on the SDSS, such as for the NGVS \citep{Raichoor2014}.

We use the meta-catalog of X-ray detected clusters (MCXC) from \citet{Piffaretti2011} to identify X--ray counterparts of our NGVS detections. This catalog has been built collecting data coming from different surveys (the ROSAT All Sky Survey and seven serendipitous surveys) and includes 1743 galaxy clusters up to $z=1.3$ 
($\langle z \rangle=0.18$). It also provides a cluster mass measurement, $M_{500}$, estimated from the 
X--ray luminosity, $L_{500}$. The catalog mass range is $10^{13}\cdot \rm M_{\odot}  \lesssim  M_{500} \lesssim 2.2 \cdot 10^{15}\ \rm M_{\odot}$, with a median of  $M_{500} = 1.8\cdot 10^{14}\ \rm M_{\odot}$.

To test the impact of the missing $r$--band observations, we use the X--ray detected group/cluster catalog from \citet{Gozaliasl2014}. It provides 135 detections in $\sim 3\ \rm deg^2$ in the CFHT--LS W1  field up to $z=1.1$. The mass range spanned by the \citet{Gozaliasl2014} catalog is $9.5\cdot 10^{12}<M_{200}<3.8\cdot 10^{14}\ \rm M_{\odot}$, with a median mass of $M_{200}=5.9\cdot 10^{13}\ \rm M_{\odot}$.

\section{The photometric redshift catalog} \label{sec:photoz}

We use the NGVS photometric redshift catalog for background sources presented in \citet{Raichoor2014}. 

 \citet{Raichoor2014} computed and compared photometric redshift estimates using the bayesian codes {\itshape LePhare} \citep{Arnouts1999,Arnouts2002,Ilbert2006} and {\itshape BPZ} \citep[][]{Benitez2000,Benitez2004,Coe2006}, obtaining similar results. In this work, we use photometric redshifts computed with {\itshape LePhare}. The adopted PSF homogenization method significantly increases the accuracy of photometric redshifts \citep{Hildebrandt2012}. 

To perform the Spectral Energy Distribution (SED) fitting, \citet{Raichoor2014} used a set of 60 templates \citep{Capak2004}, 
built interpolating four empirical galaxy spectra (Ell, Sbc, Scd, Im) \citep{Coleman1980}
and two starburst models \citep{Kinney1996}. The
reddening has been included as a free parameter ($0 < E(B -V )< 0.25$) for late type galaxies, 
applying the Small Magellanic Cloud (SMC) extinction law \citep{Prevot1984}. 
\citet{Raichoor2014} introduced a new prior  for the brightest objects: in fact, since {\itshape LePhare} has been originally built to study high-redshift objects, it has not been calibrated on the observed brightest sources ($i<20$ mag).
For the NGVS data, the photometric redshifts of low-redshift sources (z<0.2) are very important since they represent a relevant fraction of the entire sample. For this reason, the introduction of this new prior is a key point to reduce the contamination from the Virgo cluster members, which could deeply affect our analysis, and allows us to obtain more accurate photometric redshift estimations of the brightest galaxies, significantly reducing the bias, the scatter and the outlier fraction (see below for the definitions), as shown in \citet{Raichoor2014}.

To estimate the accuracy of their redshift estimates, \citet{Raichoor2014} used spectroscopic redshifts from different surveys (see their Table~4). For the NGVS fields, they used spectroscopic redshifts from the SDSS \citep{Eisenstein2001, Strauss2002, Dawson2013}, the Virgo Dwarf Globular Cluster Survey (Guhathakurta et al. 2016, {\it in preparation}), two spectroscopic programs at the {\it Anglo-Australian Telescope} (AAT)\citep[][Zhang et al. 2016, {\it in preparation}]{Zhang2015} and at the {\it Multiple Mirror Telescope} (MMT, Peng et al. 2016, {\it in preparation}).  For the CFHT--LS,  \citet{Raichoor2014} used spectroscopic redshifts from the SDSS \citep{Strauss2002}, DEEP2 \citep{cooper2008} and VVDS \citep{lefevre2013}. 

The photometric redshift uncertainty increases with magnitude and redshift  \citep{Raichoor2014}. In areas covered by five bandpasses at the depth of the CFHT-LS W1, the photometric redshift bias\footnote{the bias is defined as the median of $\Delta z=\frac{z_{phot}-z_{spec}}{1+z_{spec}}$} is $|\Delta z| < 0.02$, the scatter is $\sigma_{outl. rej}\sim 0.03 \times (1+z)$  and the fraction of outliers \footnote{Following, \citet{Raichoor2014}, outliers are defined 
 as galaxies with $|{\frac{z_{phot}-z_{spec}}{1+z_{spec}}}|>0.15$} is less than $ 9\%$.

As already discussed in the previous section, the NGVS is only partially covered by the $r$--band observations. As shown by \citet{Raichoor2014}, this affects the photometric redshift accuracy. In fact, when using the $u^*, g, i, z$-bands, the lack of the $r$-band increases the uncertainties in the $0.3 \lesssim z_{\rm phot} \lesssim 0.8$ range ($-0.05 < \Delta z < -0.02$, $\sigma_{outl.rej} \sim 0.06$, and 10-15\% outliers). 

\section{The RedGOLD detection algorithm} \label{sec:technique}
We presented our detection algorithm {\itshape RedGOLD} in L16 with its performance on semi-analytic simulations and its application on the CFHT-LS W1 field. Here we briefly summarize the method and refer to L16 for further details. 

Our algorithm relies on the observational  evidence that galaxy clusters contain a large population 
of red and bright early--type galaxies (ETGs), concentrated in their inner regions and tightly distributed on the color-magnitude diagram \citep[e.g.,][]{Gladders2000}. This assumption is observed to be true for galaxy clusters up to $z\sim1.5$ \citep[e.g.,][]{Mei2009, Snyder2012, Zeimann2012, Brodwin2013, Muzzin2013, Strazzullo2013, Mei2015}.

The method consists of the detection of  red-sequence galaxy overdensities and the confirmation of a tight red-sequence on the color-magnitude relation. To reduce the contamination due to dusty red star--forming galaxies, we select passive galaxies using two pairs of filters simultaneously, corresponding to the $(U-B)$ and $(B-V)$ rest--frame colors \citep[see][L16]{Larson1978}. 
We use \citet{Bruzual2003} (BC03) stellar population models  to compute predicted colors through the 
theoretical SEDs: we assume a single burst model, with passive evolution since the galaxy formation redshift $z_{form}=3$  and a solar metallicity, $Z= 0.02$.
In addition to this color selection, we impose that red galaxies are classified as ETGs according to the spectral classification given by {\itshape LePhare}, to select objects with spectral characteristics typical of early-type galaxies.

We define our cluster detections identifying structures with a high density contrast with respect to
the mean value of the background, estimated in each MegaCam pointing.  To retain our detections,
we also impose a constraint on the radial distribution of the red--sequence galaxies, assuming a Navarro--Frenk--White \citep[NFW;][]{Navarro1996} surface density profile.

We center our detections on a bright red ETG considering the galaxy with the highest number of red companions, weighted on luminosity.
 This approach is compatible with previous analysis, showing that centroids do not accurately trace the cluster centers, while the brightest cluster members lying near the X--ray centroid are better tracers of the cluster centers \citep{George2011,George2012}. To confirm our red overdensity based cluster candidates, we fit the red--sequence and impose limits on the red--sequence slope and scatter from \citet{Mei2009}. Finally, we assign the cluster candidate redshift considering the median photometric redshift of the passive ETGs.

To clean our cluster candidate catalog of multiple detections, we developed  an algorithm that 
identifies detections at a redshift difference $\Delta z\le 0.1$ and with at least half of the members in common. 
In the case that a multiple detection is found, we retain only the detection with the highest signal--to--noise ratio, weighted on luminosity.

\subsection{RedGOLD calibration} 

The two parameters that characterize {\itshape RedGOLD} are the cluster candidate detection significance $\sigma_{det}$ and its richness $\lambda$. 
The detection significance is defined as $\sigma_{det}=\frac{N_{gal}-N_{bkg}}{\sigma_{bkg}}$, where $N_{gal}$ is the number of red ETGs in the cell used to detect spatial overdensities, $N_{bkg}$ and ${\sigma_{bkg}}$ are the mode and the standard deviation of the galaxy count distribution in cells of the same area (i.e. represent the background contribution) and as a function of redshift.
Our richness $\lambda$ quantifies the number of bright red ETGs of our cluster candidates, using an iterative algorithm. 
In particular, we count the number of red ETGs brighter than $0.2\times L^*$ in a given radius, subtracting the scaled background.
At the zero-th iteration, we fix the scaling radius to $R_{scale}=1\ \rm Mpc$ and we estimate the corresponding richness. Successively, in the higher order iterations, we scale the
radius using the relation $R_{scale}=(\lambda/100)^{0.2}$, following \citet{Rykoff2014} (see  L16). We iterate this process until
$\Delta \lambda=\lambda_{n-1}-\lambda_n$ is comparable to the background contribution.

In L16, we calibrated the values of $\sigma_{det}$ and $\lambda$ that maximize the completeness and purity of our cluster catalog, using X--ray observations for \citet{Gozaliasl2014} and \citet{Mehrtens2012} in the CFHT-LS W1 field \citep{Gwyn2012, Erben2013} and simulations \citep{Springel2005, Guo2011, Henriques2012}. {\itshape Completeness}  is defined as the ratio of detected structures corresponding to true clusters to the total number of true clusters while {\itshape purity} is the number of detections that correspond to real structures to the total number of detected objects. For both quantities, it is important to define what is a {\it true cluster}. Following the literature  \citep[e.g, ][L16]{Finoguenov2003, Lin2004, Evrard2008, Finoguenov2009, McGee2009, Mead2010, George2011, ChiangOverzier2013, Gillis2013, Shankar2013}, we define a true cluster  as a dark matter halo more massive than $10^{14}\ \rm M_{\odot} $. Numerical simulations show that 90$\%$  of the dark matter haloes more massive than $10^{14}\ M_{\odot}$ are a very regular virialized cluster population up to redshift $z\sim1.5$  \citep[e.g.,][]{Evrard2008, ChiangOverzier2013}. 

In the fields covered by five bandpasses (see the following sections), the NGVS reaches the same photometric depth, with the same instrument and telescope, as the CFTH-LS W1, and we can use the same {\itshape RedGOLD} calibration (e.g., the same limits on the $\sigma_{det}$ and $\lambda$) obtained for the CFTH-LS W1 in  L16. In L16, we have demonstrated that, when applying {\itshape RedGOLD}  to the CFTH-LS W1 with  $\sigma_{det}\ge 4$ and  $\lambda \ge10$ at $z\le 0.6$, and $\sigma_{det}\ge 4.5$ and $\lambda \ge10$ at higher redshift,  {\itshape RedGOLD} effectively detects galaxy clusters with mass $M\gtrsim10^{14}\ M_{\odot}$, with a completeness of  $\sim100\%$($\sim70\%$) and a purity of $\sim80\%$ at $z\le 0.6$($z\lesssim1$).  Our centering algorithm and our determination of the cluster photometric redshift are very precise, with a median separation between the peak of the X--ray emission and our  {\itshape RedGOLD} cluster centers of $17.2''\pm11.2''$, and the redshift difference with spectroscopy less than 0.05 up to $z \sim 1$.

\section{NGVS galaxy cluster candidate detections} \label{sec:NGVSdetections}

We apply {\itshape RedGOLD} to the NGVS, and detect cluster candidates imposing the limits in $\sigma_{det}$ and $\lambda$ defined in the previous section. Further details on the procedure described in this section can be found in L16.

To detect spatial overdensities, we consider redshift slices of $\delta z=0.2$ in the range $0<z<1.2$,  overlapping by $\sim 3 \times \sigma_{outl. rej}$. We consider only galaxies with $i\le 23.5$~mag to detect galaxy overdensities. In fact, at fainter magnitudes, both the redshift and
photometric uncertainties are large \citep{Raichoor2014}.

Following \citet{Raichoor2014}, we identify  stars as objects with the SExtractor parameter $CLASS\_STAR>0.95$ and $i<22.5$ mag, and we remove these sources from our selection. 
As shown in \citet{Raichoor2014}, this selection removes more than 85\% of  stars while only $\sim5\%$ of galaxies are classified as stars.

  In the NGVS data, for each science image, a mask flag regions with less accurate photometry \citep[e.g. because of star haloes and the presence of extended Virgo galaxies, e.g., ][]{Erben2013, Raichoor2014}. As pointed out in \citet{Rykoff2014} (see also our discussion in L16), these masks have to be taken into account  so that the cluster richness is not underestimated. While  \citet{Rykoff2014} proposed a technique to extrapolate the richness measurement in regions with missing photometry (e.g. empty regions/holes), in L16 we chose not to use an extrapolation technique in the CFHT-LS W1. We decided to take into account the presence of masks for the stars and
other saturated objects by selecting only objects with an error in
photometry within the average distribution. In this paper, we follow the same approach, with the difference that we will also exclude areas with extended Virgo galaxies. In practice, the area over which the NGVS catalog is empty is small ($\sim10\%$ over $\sim 20\%$ of the average masked area) and the main difference in the photometry of galaxies in masked areas is the larger photometry uncertainties. We build a photometry uncertainty distribution in magnitude
bins using \citet{Raichoor2014} photometry and photometric errors, and we discard all objects that
have uncertainties more than 3-$\sigma$, which is the the average uncertainty
distribution in the red overdensity estimation.

We added masked regions to Millenium simulations to understand the bias due to masking. We included both masked regions without any source detections (e.g. empty regions/holes), and masked regions with higher photometry uncertainties. When running {\itshape RedGOLD}  on the masked modified Millennium simulations, the recovered purity and completeness levels do not differ from those obtained without considering the masked regions.
We also check that masks do not significantly change our definition of richness. For each detected cluster candidate, we estimate the richness $\lambda_{mask}$, including also sources that are not included in our richness estimate because have large photometric errors in the \citet{Raichoor2014} NGVS photometric catalog. 

As for the the CFHT-LS W1, $\sim 7\%$ of the {\it RedGOLD} cluster candidates (obtained without imposing our lower limits on $\lambda$, $\sigma_{det}$ and the radial galaxy distribution) have a fraction of masked bright potential cluster members $>10\%$. If we consider only the {\it RedGOLD} detections obtained imposing our lower limits, we find that   $\sim 2$\%  have a fraction of masked bright potential cluster members $>10\%$. As in L16, we conclude that our richness estimate is not significantly affected by the presence of the masks for at least $\sim 98$\% of the cluster candidates. We also examined in which objects the fraction of masked members impacts our richness measurements, and we obtain that we can estimate richness $\sim 10\%$ lower, for partially masked clusters with low richness and high redshift ($\overline \lambda_{mask}=12$, $\overline z_{cluster}=0.7$). 

Also, the fact that we do not consider NGVS areas with holes and high photometric uncertainties means that we do not detect clusters in the areas where extended (bright) Virgo galaxies are masked. Our completeness is estimated in unmasked regions.

As shown in \citet{Raichoor2014}, though the global photometric redshift accuracy remains high even when using only four optical bands, the uncertainty on the photometric redshifts for sources at $0.3<z<0.8$ is larger, because 
the $r$-band samples the $4000$ \AA\ break in this redshift range (see section \ref{sec:photoz}). 

Among the fields covered by the $r$--band observations, eleven co-added images only consist of one or two individual exposures in the $r$--band. This leads to an incomplete and very inhomogeneous coverage of the MegaCam field of view,  and  implies that in those fields the data quality in the $r$--band is lower, because of both the lower depth, and the lack of coverage of the intra-CCD regions, due to the lack of an adequate number of dithered 
exposures.

The former has an impact on the detection of the less massive structures at intermediate and high redshifts, where the shallower data prevent the detection of the fainter galaxies on the red--sequence. As a consequence, the red-sequence appears to be less populated than expected, and the contrast of the red--sequence cluster galaxies with respect to the background is lower. 
The latter  does not significantly affect the efficiency of our algorithm in the detection of galaxy overdensities since our detections are based on the contrast relative to the background, but it has a strong effect when estimating the cluster centers and their richness.

In fact, if part of a cluster is masked,  we might be able to detect it because the contrast with respect to the background will still be significant, but we would obtain less accurate cluster centers if the inner region of the cluster falls in the masked area. As a consequence, the richness would be deeply underestimated because of both the cluster miscentering \citep[e.g.,][]{Johnston2007,Rozo2011} and the missing galaxies
that are in the masked regions. Since we iteratively estimate $\lambda$ in the scaling radius, $R_{scale}$ will soon become too small (typically of the order of 200-300 kpc) to 
give reliable richness estimates. 

For this reason, in the eleven fields covered by shallower $r$--band observations, we do not use the colors based on the r--filter when searching for red galaxy overdensities and estimating the cluster center and richness, but we use different bands considering these fields in the same way as the fields that are only covered by four bandpasses. 
At $z\le0.5$ we use the color pairs $(g-i)$ and $(g-z)$ while at $z>0.6$ we impose only one color limit, $(i-z)$. 

\subsection{Differences in the richness estimates for fields without $r$-band observations}\label{sec:richcomp}

The richness $\lambda$ is one of the two parameters that we optimize to obtain our estimations of completeness and purity (L16). We expect that the lack of r--band impacts our estimation of $\lambda$, and we perform a simple test to quantify the typical difference between the cluster richness estimated with a full band coverage (hereafter, $\lambda_r$), and without the $r$--band (hereafter, $\lambda_{wr}$). 

We compare the two richness estimates in the 23 fields with deep $r$--band observations, and in  Fig. \ref{fig:CompRich} we plot  the histogram of $\Delta \lambda/\lambda_r=(\lambda_r-\lambda_{wr})/\lambda_r$ in different redshift bins. 

Fig.~\ref{fig:richZ} shows $(\lambda_r-\lambda_{wr})/\lambda_r$ as a function of redshift. The red dashed line represents  $(\lambda_r-\lambda_{wr})/\lambda_r=0$. Table \ref{tab:lambda} shows the median value of $(\lambda_r-\lambda_{wr})/\lambda_r $ and its standard deviation $\sigma_{\Delta \lambda}$ in redshift bins. 
At redshift $z<0.5$, the two richness estimates are in good agreement ($\Delta\lambda/\lambda_r <10\%$).  In fact, the difference between the two measurements is not significant because when we use the color $(g-z)$ instead of $(g-r)$ to select red sequence galaxies, it still straddles the 4000 \AA\ break. 
At $0.5<z<0.6$, the richness estimated without the $r$--band data is systematically underestimated by $\sim 40\%$ on average. This is because we use the $(g-i)$ color, which changes less steeply as a function of redshift than the $(r-i)$ and $(i-z)$ colors, and has larger photometric errors.
At higher redshifts, $0.6<z<0.8$, we systematically over--estimate  the optical richness estimated without the $r$--band by $\sim 20\%$, on average. In fact, when the r--data are unavailable, we only use the $(i-z)$ color constraint to identify red cluster members, while, when also using r--observations, we consider an additional cut in the $(r-z)$ color ($(r-i)$ at $z\sim0.6$), which contributes to the reduction of the contamination of dusty red galaxies on the red-sequence. 
Finally, at $z>0.8$, the two richness estimates are in agreement because the $(i-z)$ color straddles the $4000$ \AA\ break and it successfully isolates passive cluster members.

Since this comparison shows that at $z<0.5$ and $z\ge 0.8$, $\lambda_{wr}$ is on average  under-- or over--estimated by a factor $<10\%$, in this redshift range, we do not adopt a different richness threshold in the cluster catalog built using only four optical bands (hereafter, {\it  RedGOLD$_{wr}$})  in these redshift intervals. However, at $0.5\le z<0.8$, $\lambda_{wr}$ is on average significantly under-- or overestimated (by a factor of $\sim20-40\%$, depending on the redshift bin; see Table \ref{tab:lambda}). 
This means that at $0.5<z<0.6$,  we could discard real cluster candidates with $1\lesssim M_{200}\lesssim2 \times 10^{14}\ M_{\odot}$, and at $0.6\le z <0.8$, we could include more false detections. 

For this reason, in these redshift ranges, we apply different richness thresholds, estimated as $\lambda_{wr}\ge  \lambda_{r, min}+ \lambda_{r, min}\times median( \Delta\lambda)$, where $\lambda_{r, min}=10$ is the adopted optimized threshold when using the full band coverage.
As we will discuss in section~\ref{sec:4band}, this choice allows us to obtain both completeness and purity comparable with the cluster candidate catalog built using the full band coverage (hereafter, {\it  RedGOLD$_{r}$}). 
The observed difference in richness should be carefully taken into account when studying scaling relations, such as the mass-richness, because it could introduce larger scatters.

\subsection{Cluster catalog with deep $r$-band observations} \label{sec:deep}

In the  $\sim 20\ \rm deg^2$ covered with deep $r$-band data, when (without) applying the thresholds on the {\it RedGOLD} parameters, we detect 294 (1045) cluster candidates, i.e. $\sim 15 $ detections per square degree when applying the thresholds, in agreement with the values predicted for our cosmological model \citep{Weinberg2013} in the same mass range. 
The $57\%$ ($31\%$) of the cluster candidates detected with (without) thresholds,
have at least one SDSS spectroscopic member in less than $1.5'$ with $|z_{spec}-z_{cluster}|<0.1$. 

In the $\sim 20\ \rm deg^2$ covered by deep $r$-band observations,  there are four X--ray detected clusters in the \citet{Piffaretti2011} catalog at redshift $0.05<z<0.25$. {\it RedGOLD$_r$}
 is able to recover three of them (either or not imposing the considered thresholds on the {\it RedGOLD} parameters), with masses included in the range $1.4 \cdot 10^{14} \rm M_{\odot}<M_{500}<4.2\cdot10^{14} \rm M_{\odot}$ and $0.08<z<0.2$. We miss one X--ray detection at $z=0.067$, with an estimated  mass of $M_{500}=2.7\times 10^{13} \rm M_{\odot}$. This is not surprising, since we expect a completeness of $\lesssim 50\%$ in this mass and redshift range (L16).
 
 In Fig.~\ref{fig:spatDistrNGVS}, we show the spatial distribution of our detections in the fields covered by deep $r$--band observations as red points. In  Fig.~\ref{fig:CompDistrRedsMeRedM} and Fig.~\ref{fig:CompDistrRichMeRedMDeep}, we  show their redshift and richness distribution (red solid line), respectively. 
For comparison, in  Fig.~\ref{fig:CompDistrRedsMeRedM} and  Fig.~\ref{fig:CompDistrRichMeRedMDeep}, we plot the same distributions for the redMaPPer detections in the same area (black dashed line). The histograms are normalized to the corresponding total number of detections.

To match the {\it RedGOLD} cluster candidates with the redMaPPer catalog, we adopt the same matching algorithm described in L16, with
 a maximum projected distance between the centers corresponding to $R_{200}+\sigma_{R200}$ and a maximum
redshift difference of  $\Delta z=|z_{redMaPPer}-z_{RedGOLD}|\le3\times \sigma_{photoz}=3\times 0.03\times(1+z_{RedGOLD})$, where $z_{redMaPPer}$ is the cluster redshift in the redMaPPer catalog. 

In the $\sim 20\ \rm deg^2$ covered by deep $r$--band observations, {\itshape RedGOLD$_r$} recovers all the 53 redMaPPer clusters,  without limits on $\sigma_{det}$, $\lambda$, and the cluster radial profile. When applying limits in $\sigma_{det}$, $\lambda$ and the cluster radial profile, we obtain 46 clusters, for a recovery of $87^{+4}_{-6}\%$/$100\%$, when using/not using the thresholds. The seven detections that are discarded when using the thresholds have significance levels below our adopted threshold in $\sigma_{det}$, and/or richness $\lambda_r<10$.

 In Fig.~\ref{fig:CompRich} and  Fig~\ref{fig:CompRichHisto}, we compare the richness estimates obtained by redMaPPer and {\itshape RedGOLD} for the 46 common detections. We show  
$\lambda_{r}$ vs $\lambda_{redMaPPer}$ and the histogram of the difference between our richness definition and the richness adopted in \citet{Rykoff2014}. 
Different colors show the observed difference in different redshift bins, as indicated in each panel. In the bottom right panel in Fig.~\ref{fig:CompRichHisto}, we plot the ($\lambda_{redMaPPer}-\lambda_{r})/\lambda_{r}$ as a function of redshift: as already shown in L16, the difference between the two richness estimates in the {\it RedGOLD} and redMaPPer catalog is larger at higher redshift. We have shown that for our CFHT-LS W1 {\itshape RedGOLD} cluster candidate sample, the redMaPPer richness is systematically higher than the {\itshape RedGOLD} richness at $z>0.3$, and Table~4 of that work gives the median ($\lambda_{redMaPPer}-\lambda_{r})/\lambda_{r}$ in five different redshift bins. For this work, in the $\sim 20\ \rm deg^2$ covered by deep $r$--band observations, we do not have enough statistics to obtain solid median difference estimates as in L16. In Table~\ref{Tab:CompRichr}, we present the median value of this richness difference as a function of redshift.  These results are consistent with Table~4 in L16 and show that the median difference is small at low redshift  ($z<0.3$), but  increases  at higher redshifts. The first bin has few detections and it is not statistically significant. 

The {\it RedGOLD$_r$} catalog  includes 150 new detections 
 with respect to the redMaPPer catalog in the same area and the redshift range covered by the redMaPPer catalog ($z\leq 0.55$). We extensively discussed the comparison between  {\it RedGOLD$_r$}--like and redMaPPer detection in section 6.3 in L16, where we concluded that because of the richness limit imposed to the redMaPPer catalog (Rozo, private communication)  and the greater depth of the NGVS compared to the SDSS, {\it RedGOLD$_r$} detects cluster candidates at lower mass and higher redshift than those in the redMaPPer published catalog (as it is shown in Fig.~\ref{fig:CompDistrRedsMeRedM} and  Fig.~\ref{fig:CompDistrRichMeRedMDeep}). A more detailed comparison of the two detection algorithms should be done on exactly the same photometric and photometric redshift catalogs on the same area, and this is a good subject for future work.

\subsection{Cluster catalog with  four bandpasses}\label{sec:4band}
To build a homogenous selection on the whole NGVS area, we also provide a cluster catalog using only four bandpasses, adopting the same approach described in the case of the fields covered by deep $r$-band observations, but isolating red-sequence galaxies using different color pairs, as described in section \ref{sec:NGVSdetections} and using a corrected $\lambda_{wr}$ limit, as explained in section~5.1. 

 When (without) applying the thresholds on the {\it RedGOLD} parameters, {{\it RedGOLD$_{wr}$} finds  1724 (6233) cluster candidates up to $z=1.1$, i.e. $\sim15$ detections per square degree when applying the thresholds. The $\sim62\%$ ($\sim36\%$) of the cluster candidates detected with (without) the thresholds have at least one SDSS spectroscopic member in less than $1.5'$ with $|z_{spec}-z_{cluster}|<0.1$.

To quantify the completeness and the purity of  the {\it RedGOLD$_{wr}$} catalog and compare it with {\it RedGOLD$_{r}$}, we apply {\it RedGOLD$_{wr}$} to $\sim3\ \rm deg^2$ in the CFHT-LS W1 field, covered by the X--ray detected cluster catalog by \citet{Gozaliasl2014}, which includes 135 clusters and groups (hereafter, the CFHT-LS W1 GZ field).

As discussed in L16, in the CFHT-LS W1 GZ field, \it  RedGOLD$_{r}$} finds 38 cluster candidates up to $z\sim1$ and 28 of them have an X--ray counterpart in  \citet{Gozaliasl2014}. Among the \citet{Gozaliasl2014} clusters with $ M_{200}>10^{14}\ \rm M_{\odot}$, {\it  RedGOLD$_{r}$} recovers 100\% of them up to $z=0.5$ and $76^{+11}_{-15} \%$ in the whole redshift range.  In the same area, {\it RedGOLD$_{wr}$} finds 42 cluster candidates up to $z\sim1$, of which 33 are in common with \citet{Gozaliasl2014} and 30 are also detected by  {\it  RedGOLD$_{r}$}. Among the \citet{Gozaliasl2014} clusters with $ M_{200}>10^{14}\ \rm M_{\odot}$, {\it  RedGOLD$_{wr}$} recovers all of them up to $z=0.5$ and $71^{+12}_{-15}\%$ in the whole redshift range. Two cluster candidates detected by  {\it RedGOLD$_{wr}$} and without an X--ray counterpart have been spectroscopically confirmed as clusters at $z=0.33$ \citep{Andreon2004} and $z=0.92$ \citep[][C3 cluster]{Pierre2006}. This implies that the lower limit on the purity with {\it RedGOLD$_{wr}$}  is of $\sim 80\%$.

 This means that, when adopting different thresholds of $\lambda$ as a function of redshift according to the median values reported in Table~\ref{tab:lambda}, we obtain similar values of completeness and purity with  {\it  RedGOLD$_{r}$} and {\it  RedGOLD$_{wr}$}.

To directly compare the {\it RedGOLD$_{wr}$} and {\it RedGOLD$_{r}$} catalogs, we also checked the detections that are not found in both cases. 

Eight {\it RedGOLD$_{r}$} candidates are not detected by {\it RedGOLD$_{wr}$}: two are discarded  when performing the fit of the color-magnitude relation,  four because they have $\sigma_{det}<4$, one because of the galaxy distribution constraint, and one when imposing our lower limit on the richness.
Twelve {\it RedGOLD$_{wr}$} cluster candidates are not detected with {\it RedGOLD$_{r}$}. Six of them have an X--ray counterpart in the group catalog by \citep[][$M\le8.5\times 10^{13}\ \rm M_{\odot}$]{Gozaliasl2014} and one is a spectroscopically confirmed system at $z=0.92$ \cite{Pierre2006}. The remaining five {\it RedGOLD$_{wr}$} detections are detected as red--overdensities also by {\it RedGOLD$_{r}$}, but they are discarded because of the imposed constraints: in particular, one is discarded when performing the fit of the CMR, two  are discarded because of our constraints on the NFW profile, one because of its detection significance ($\sigma_{det}<4.5$) and  one because of its richness estimate $\lambda_r$. 

These results suggest that the additional cluster candidates detected by {\it RedGOLD$_{wr}$} and unrecovered by {\it 
RedGOLD$_{r}$}  correspond to less massive systems.  

To summarize, the lack of the $r$--band seems to have a minor effect on the 
{\it RedGOLD} completeness and does not impact its purity.
With {\it RedGOLD$_{wr}$}, we miss only $\sim5\%$ of the {\it RedGOLD$_{r}$} detections more massive than 
$10^{14}\ \rm M_{\odot}$ at higher redshift (z>0.5), while we obtain the same completeness  of $100\%$ at lower redshift.

In Fig.~\ref{fig:spatDistrNGVS4b}, we show the spatial distribution of the {\it RedGOLD$_{wr}$} detections. In Fig.~\ref{fig:redsDistrNGVS4b}, we show their redshift distribution (solid line).

 For comparison, in Fig.~\ref{fig:redsDistrNGVS4b}, we show the redMaPPer detections (dashed line) in the same area. Up to $z\sim0.55$,  we find that {\it RedGOLD$_{wr}$} recovers all the 267 redMaPPer candidates.
When applying limits in $\sigma_{det}$, $\lambda$ and the cluster radial profile, we obtain 238 detections, for a final recovery of $90^{+2}_{-2}\%$/$100\%$ with/without thresholds in the  {\it RedGOLD} parameters. 

In Fig.~\ref{fig:CompDistrRichMeRedMNoSDSS}, we compare the richness distribution of the {\it RedGOLD$_{wr}$} cluster candidates (solid line) with respect to the richness of the redMaPPer detections recovered by {\it RedGOLD$_{wr}$} (dashed line). 
In the {\it RedGOLD$_{wr}$} catalog, we detect 1032 new cluster candidates in the same area and redshift range of the redMaPPer catalog ($z \leq 0.55$), which are distributed at lower richness with respect to the redMaPPer catalog (i.e. $\sim 10$ new cluster candidate per $\rm deg^2$ up to $z\sim0.55$), because of both the richness threshold adopted in the redMaPPer catalog (Rozo, private communication) and the higher depth of the NGVS with respect to the SDSS data.

 In Fig.~\ref{fig:CompRich} and  Fig~\ref{fig:CompRichHisto} we compare the richness estimates obtained by redMaPPer and {\itshape RedGOLD} for the 238 common detections. We show  the 
$\lambda_{wr}$ vs $\lambda_{redMaPPer}$ and the histogram of the difference between our richness definition and the richness adopted in \citet{Rykoff2014}. 
Different colors show the observed difference in different redshift bins, as indicated in each panel.
Consistently with our results in L16, the  redMaPPer richness is systematically higher than the {\itshape RedGOLD} richness at $z>0.3$. In the bottom right panel in Fig.~\ref{fig:CompRichHisto}, we plot the ($\lambda_{redMaPPer}-\lambda_{wr})/\lambda_{wr}$ as a function of redshift: as already shown in L16, the difference between the two richness estimates in the {\it RedGOLD} and redMaPPer catalog is larger at higher redshift.
In Table~\ref{Tab:CompRich}, we present the median value of this richness difference as a function of redshift.  The median difference is small at low redshift ($z<0.3$), but it increases at higher redshifts.

These results are consistent with our results in the fields covered by five bands (Table~\ref{Tab:CompRichr}) and L16, within the errors. As discussed in L16, we believe that this difference might be due to the fact that we keep a simple approach counting galaxies up to the depth reached by the CFHTLenS, while the redMaPPer richness estimate includes an extrapolation of the SDSS depth (which is lower than CFHTLenS) to our same limit in ${\rm L^*}$.  It would be worth investigating the observed difference richness in a future work in collaboration with the redMapper authors, considering a larger cluster sample and using the same photometric and photometric redshift catalogs.

Finally, we compare the {\it RedGOLD$_{wr}$} detections  with the nine X--ray cluster catalog by \citet{Piffaretti2011}: we find that {\it RedGOLD$_{wr}$} is able to recover seven of them without imposing constraints on the $\sigma_{det}$, $\lambda$ and the cluster radial profile, in the range $1.3  \cdot 10^{14} \rm M_{\odot}<M_{500}<4.3  \cdot 10^{14} \rm M_{\odot}$ and $0.08<z<0.55$. Three of them  are also detected by {\it RedGOLD$_{r}$}. The other two clusters have masses of $M_{500}=3-5  \cdot 10^{13}$ (they are low mass systems) and $z=0.04-0.09$. 
When considering our lower limits on the {\it RedGOLD} parameters, we discard two of the detected clusters because of their $\sigma_{det}$ or the radial galaxy distribution. The clusters that are discarded while imposing limits have masses $M_{500} \sim 1.4 \cdot 10^{14}\ \rm M_{\odot}$ and $z\sim0.08$.

\subsection{The NGVS Cluster Candidate Catalogs}
 The NGVS {\it RedGOLD} cluster candidate catalogs, obtained with the thresholds on the  {\it RedGOLD} parameters, will be public when this paper is published.
We provide two independent catalogs: one for the $\sim20\ \rm deg^2$ with deep $r$--band observations, and the other for the whole NGVS area, using only four optical bands.
 
In our catalogs, we include the following parameters for each {\it RedGOLD} detection:
\begin{enumerate}
\item the J2000 right ascension RA 
\item the J2000 declination decl.
\item the photometric cluster redshift PHOTZ
\item the error in photometric redshift ePHOTOZ
\item the average spectroscopic redshift, when available SPECZ
\item the detection significance $\sigma_{det}$ SDET
\item the cluster richness $\lambda$  RICH
\item the uncertainty on cluster richness $\lambda_{err}$ eRICH
\end{enumerate}

Virgo galaxies and globular clusters are excluded from our sample by the photometric and photometric redshift selection, as shown in Raichoor et al. (2014). However, the NGVLens reduction includes masking areas for which holes and the noise due to bright Virgo galaxies would prevent background galaxy detection. This means that, to have a homogeneous photometry and photometric redshift catalog, we do not detect clusters in the masked areas. 
To quantify how this biases our catalogs, in Fig.~\ref{fig:Dens} we show the density distribution of our detections in each MegaCam field using four bandpasses. Different colors correspond to the different number of cluster candidates per MegaCam pointing, as shown in the color bar, and the black stars represent the  Virgo cluster members from \citet{Kim2014}. 
The symbol size is proportional to the corresponding Kron radius from \citet{Kim2014}. The white star represents M87.

When using four bandpasses, the average number of clusters per square degree is $15\pm4$ (the uncertainty is the standard deviation of the distribution of detection for each pointing), consistent with our detections in the CFHT-LS W1(L16). We obtain a corrected $19\pm 6$ when the number of detected clusters in each MegaCam pointing is divided by the unmasked area in that pointing. In Fig.~\ref{fig:Dens}, density variations in different pointings are all consistent within $\sim 2\sigma$, independently on the masks due to the presence of stars and Virgo galaxies. Also, when we perform a Pearson, Spearman and Kendall correlation test on the number of detection vs the masked area, we obtain a probability of $42-45\%$ that the two variables are not correlated. This means that statistical variations in the cluster detection distribution are more significant than the correlation between the number of detected clusters and the masked area (e.g. the presence of stars and Virgo galaxies). We deduce that the presence of bright Virgo galaxies prevent us from detecting clusters in the masked areas, and, from the percentage of masked area and assuming that the cluster distribution is the same in the masked and unmasked area, we estimate that  $\sim 20\%$ of the clusters will be missed. However, as shown in L16, this does not significantly bias our detections in the unmasked area. We have already discussed in Sec.~\ref{sec:4band} that the lack of the r-band biases our richness estimates. This should be taken into account when using these catalogs for cosmology and other predictions based on the number of detected clusters.

In Fig.~\ref{fig:cutout} we show the optical images of four {\it RedGOLD} detections in the
NGVS field.

\section{Conclusions} \label{sec:conclusions}

We apply our cluster detection algorithm {\itshape RedGOLD} (L16) to deep optical observations from the NGVS \citep{Ferrarese2012}, which covers  $104 \ \rm deg^2$ around the center of the Virgo
 cluster.  
 
 {\it RedGOLD} is a cluster detection algorithm based on the selection of passive galaxy overdensities, simultaneously using two pairs of filters, corresponding to the $(U-B)$ and $(B-V)$ rest--frame colors (L16).  It also imposes constraints on the detections to be compatible with an NFW profile.
  For each cluster candidate, it estimates the 
detection significance $\sigma_{det}$, and provides a richness measurement $\lambda$ based on the number of luminous red--sequence galaxies.
In L16, we showed that, at the NGVS depth,
the best compromise between purity and completeness is reached for detections with richness $\lambda\ge10$ and detection significance $\sigma_{det}\ge 4\ (\ge 4.5)$ at $z\le 0.6$ ($z\lesssim1$), according to both empirical calibration on the X--ray group catalog by \citet{Gozaliasl2014} and simulations \citep{Springel2005, Guo2011, Henriques2012}. With these limits, we expect our cluster candidate catalogs to be   $\sim100\%$($\sim70\%$) complete and $\sim80\%$ pure, at $z\le 0.6$($z\lesssim1$), for galaxy clusters with masses $M\gtrsim10^{14}\ M_{\odot}$.  Our cluster centering algorithm attains a  median separation between the peak of the X--ray emission and our  {\itshape RedGOLD} cluster centers of $17.2''\pm11.2''$, and our determination of the cluster photometric redshifts differ from spectroscopic redshifts by less than 0.05 up to $z \sim 1$ (L16).

 Part of the NGVS area, $\sim 20\ \rm deg^2$, has been observed in five bandpasses ($u^*,g,r,i,z$) to the nominal NGVS depth \citep{Ferrarese2012, Raichoor2014}.  Over the remaining survey area,
 the $r$--band is either shallower ($\sim 10\ \rm deg^2$; see Table~\ref{tab:NGVSdepth}), or has not been observed 
 ($\sim 74\ \rm deg^2$).
 In these fields, the photometric redshift estimates are less accurate \citep{Raichoor2014}, 
 especially in the redshift range of $0.3< z<0.8$, since the missing $r$--band is one of the two bands that 
 straddles the $4000$ \AA\ break. 
 
Because of this difference in the $r$--band coverage, we build two independent {\it RedGOLD} cluster catalogs: one using the deep $r$--band observations, available only over $\sim 20\ \rm deg^2$ (the {\it RedGOLD$_r$} catalog), and the other using only four bandpasses over the entire NGVS area (the {\it RedGOLD$_{wr}$} catalog).  

We show that the  lack of the $r$--band observations impacts the estimation of the cluster richness $\lambda$. To estimate the corresponding bias, we use the $\sim20\ \rm deg^2$ covered by deep $r$--band data and compare the two cluster richness estimates obtained with five and four bandpasses, i.e., with {\it RedGOLD$_r$} and {\it RedGOLD$_{wr}$}, respectively. 
We find that the lack of the $r$--band observations does not affect the cluster richness estimate up to $z\sim0.5$ and at $z>0.8$. 
At $0.5\le z<0.6$, the cluster richness estimated without using the $r$--band is  underestimated by a factor of $\sim 40\%$, while at $0.6\le z<0.8$ it is systematically overestimated by a factor of $\sim 20\%$. As a consequence, we adopt different lower limits on $\lambda$ when applying {\it RedGOLD$_{wr}$} at these redshifts, taking into account the median estimated difference $\lambda_r-\lambda_{wr}$. With this choice, using the X--ray group catalog by \citet{Gozaliasl2014}, we demonstrate that  {\it RedGOLD$_{r}$} reaches similar completeness and purity as {\it RedGOLD$_{wr}$}. 

Over the $\sim 20\ \rm deg^2$  with complete and deep r--band coverage, we find 294 (1045) cluster candidates up to $z\sim1$, i.e. $\sim 15$ detections per $\rm deg^2$, when applying (not applying) the threshold on {\it RedGOLD} parameters.
Of the cluster candidates detected with (without) considering lower limits on the cluster richness, the detection significance or the cluster radial distribution,  $57\%$ ($31\%$) have at least one SDSS spectroscopic member in less than $1.5'$ with $|z_{spec}-z_{cluster}|<0.1$. 

When using four optical bands, we find 1724 (6233) cluster candidates up to $z\sim1$ over the entire NGVS area, i.e. $\sim 15$ detections per $\rm deg^2$, when applying (not applying) the threshold on {\it RedGOLD} parameters.   Of the cluster candidates detected with (without) considering lower limits on the cluster richness, the detection significance or the cluster radial distribution, $62\%$ ($36\%$)  have at least one SDSS spectroscopic member in less than $1.5'$  with $|z_{spec}-z_{cluster}|<0.1$.

With {\it RedGOLD$_{wr}$}, we recover most of the X--ray detected clusters with $M_{500} > 1.4 \times 10^{14} \rm M_{\odot}$ and $0.08<z<0.2$.
{\it RedGOLD} recovers all of the redMaPPer detections \citep{Rykoff2014} in all fields when we do not apply the thresholds on {\it RedGOLD}  parameters. When we do apply our thresholds on the {\it RedGOLD}  parameters, we find  $90\%$ of redMaPPer detections over the deep $r$--band data area, and $87\%$ of redMaPPer detections over the entire NGVS area in which we used only four bands.

These results confirm that {\itshape RedGOLD} successfully detects galaxy clusters, and show that even when using only four optical bands, {\it RedGOLD} is able to provide cluster catalogs with high completeness and purity, as shown in L16.

The NGVS cluster candidate catalogs will be made public on the NGVS website once this paper appears in publication.

\acknowledgments

The French authors (R.L., S.M., A.R.) acknowledge the support of the French Agence Nationale
de la Recherche (ANR) under the reference ANR10-
BLANC-0506-01-Projet VIRAGE (PI: S.Mei). S.M. acknowledges financial support from the Institut Universitaire de France (IUF), of which she is senior member.
H.H. is supported by the DFG Emmy Noether grant Hi 1495/2-1. We thank the Observatory of Paris for hosting T.E. under its visitor program. We thank Olivier Ilbert for useful comments. We thank James G. Bartlett for the interesting discussions and for carefully editing the abstract and the conclusions.




{\it Facilities:} \facility{CFHT}

\bibliographystyle{apj}
\bibliography{myRefMajor}

\clearpage




\begin{figure*}
\epsscale{2}
        \plotone{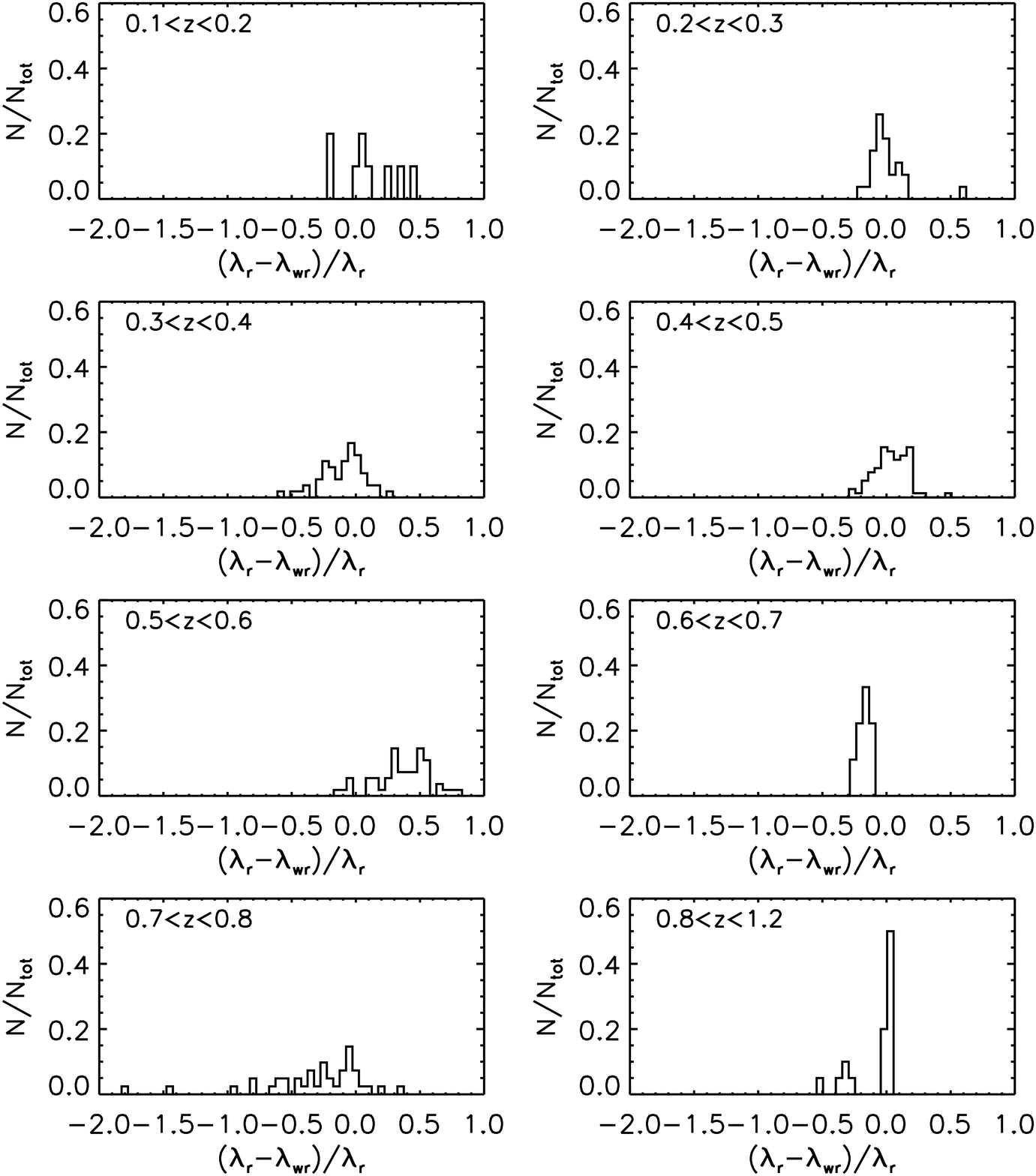}    
    \caption{Comparison of the richness estimated using {\it RedGOLD$_{r}$} (the algorithm is run using the full band coverage, five bands) and {\it RedGOLD$_{wr}$} (the algorithm is run without using the $r$--band), in the $23\ \rm deg^2$ covered by deep $r$--band observations. Each panel shows the distribution of $(\lambda_r-\lambda_{wr})/\lambda_r$ in different redshift bins, where
    $\lambda_{wr}$ is the richness estimated without using the $r$--band and $\lambda_r$ is the richness estimated using the deep r--band data, which we are assuming to be our reference cluster richness.}
        \label{fig:CompRich}
  \end{figure*}

\clearpage

\begin{figure*}
\epsscale{1}
               \plotone{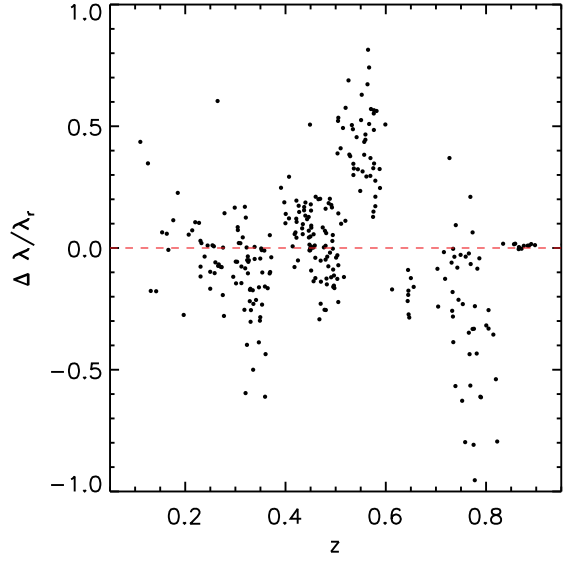}    
    \caption{Comparison of the richness estimated using estimated using {\it RedGOLD$_{r}$} and {\it RedGOLD$_{wr}$} as a function of redshift. The red dashed line represents $(\lambda_r-\lambda_{wr})/\lambda_r=0$.}
    \label{fig:richZ}
  \end{figure*}

\begin{figure*}
\epsscale{1}
               \plotone{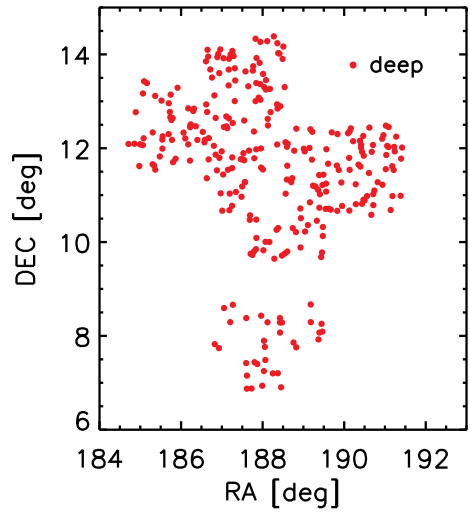}    
    \caption{Spatial distribution of the NGVS cluster detections in the $\sim 20\ \rm deg^2$ covered by deep $r$-band observations. Red points represent detections in the fields with deep  r--data, i.e. {\it RedGOLD$_{r}$} cluster candidates.}
    \label{fig:spatDistrNGVS}
  \end{figure*}

\begin{figure*}
\epsscale{1}
   \plotone{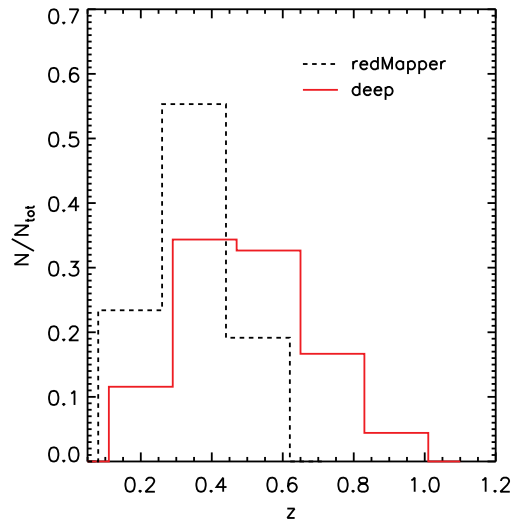}    
    \caption{The redshift distribution of the {\it RedGOLD$_{r}$}  cluster candidates (red solid line). The redMaPPer redshift distribution   is shown by the black dashed line.  Each histogram is normalized to the total number of detections.}
    \label{fig:CompDistrRedsMeRedM}
  \end{figure*}

\begin{figure*}
\epsscale{1}
       \plotone{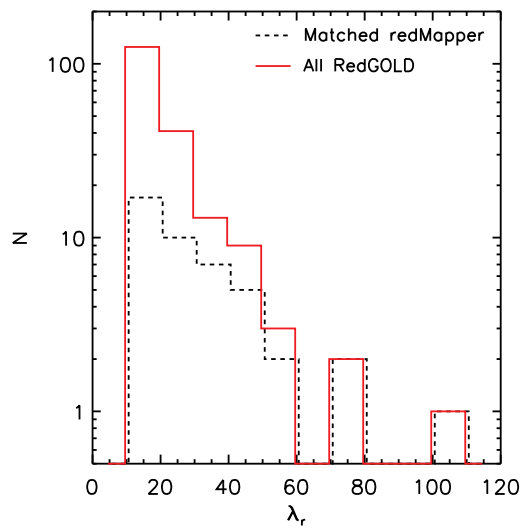} 
    \caption{Richness distribution of the 294 {\it RedGOLD$_{r}$} cluster candidates  (red solid) and of the RedGOLD candidates with a counterpart in the redMaPPer catalog (black dashed) in the $\sim 20\ \rm deg^2$ with deep $r$--band observations. Our catalog includes 150 detections (in the same area and redshift range, $z \leq 0.55$) that are not in the redMaPPer public catalog (see the text).}
    \label{fig:CompDistrRichMeRedMDeep}
  \end{figure*}
  
  \begin{figure*}
    \begin{center}
 \epsscale{1.5}
    \plotone{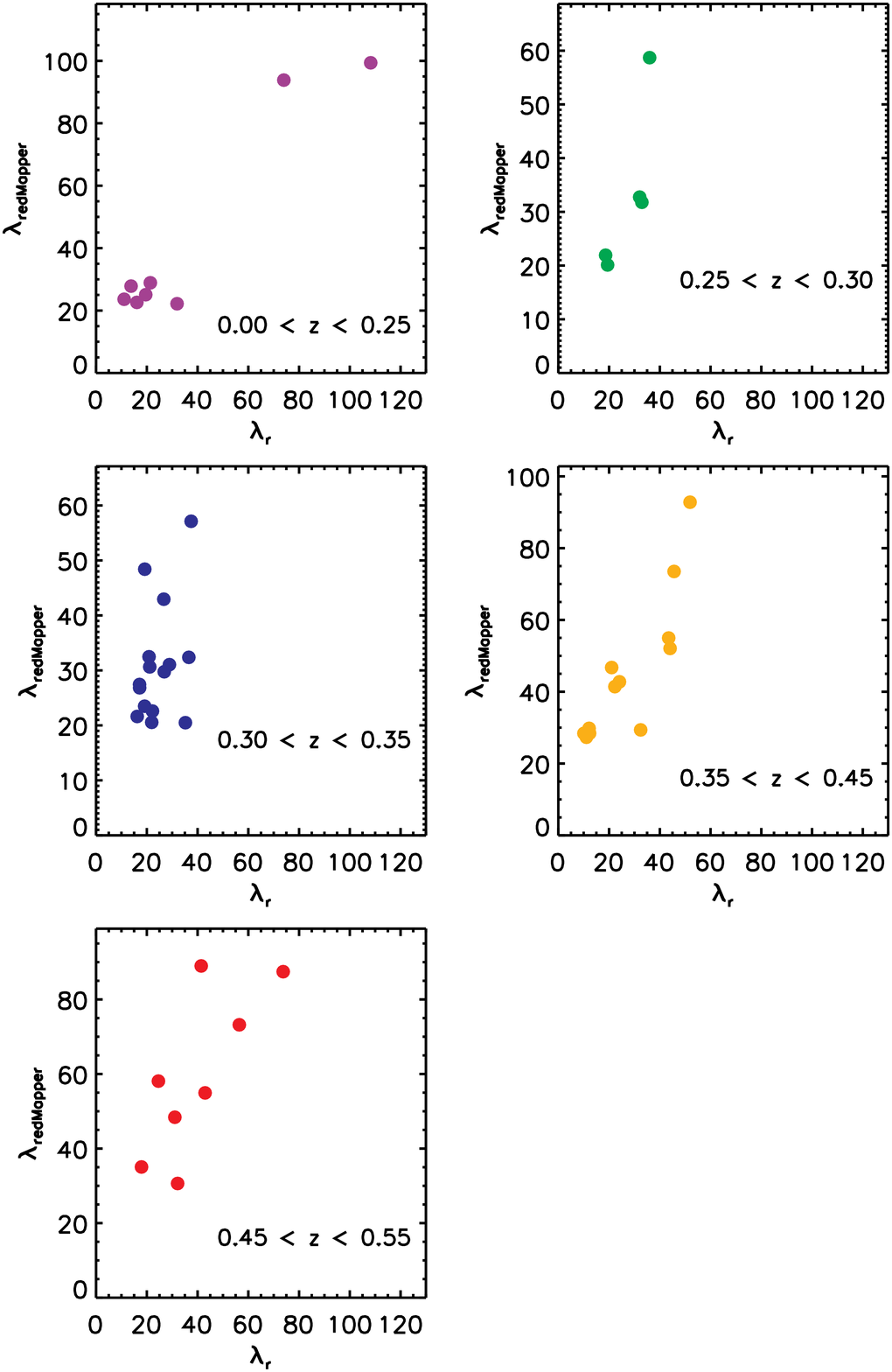}       \end{center}
    \caption{Comparison of the {\it RedGOLD$_{r}$} richness ($\lambda_{r}$) vs the redMaPPer richness ($\lambda_{redMaPPer}$)  in different redshift bins as indicated in each panel. }
    \label{fig:CompRich}
  \end{figure*}

\begin{figure*}
    \begin{center}
       \epsscale{1.5}
    \plotone{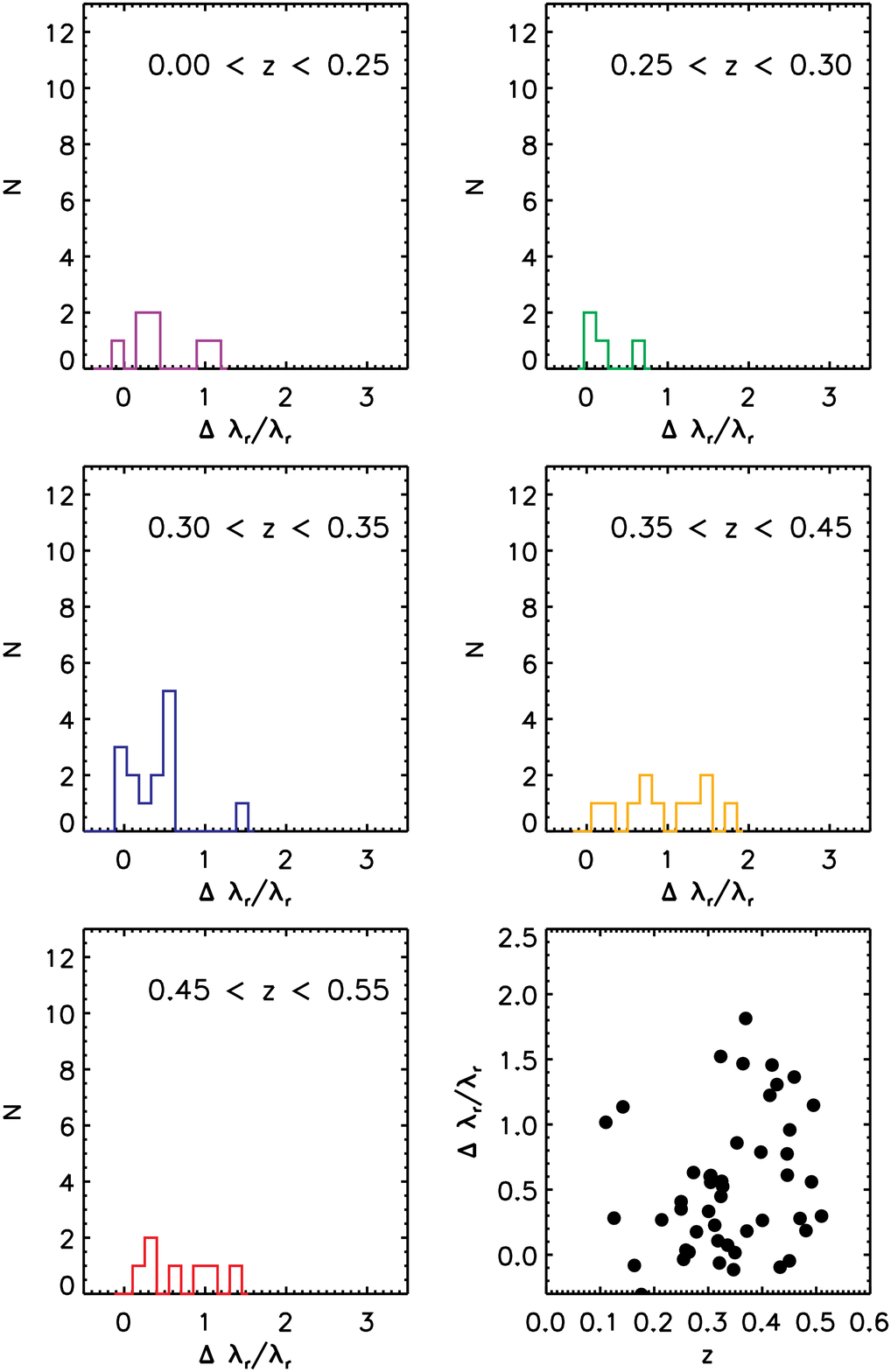}       \end{center}
    \caption{Histogram of  $\frac{\Delta \lambda_{r}}{\lambda_{r}} = \frac{\lambda_{redMaPPer}-\lambda_{r}}{\lambda_{r}}$, in different redshift bins as indicated in each panel. The bottom right panel shows the $\frac{\Delta \lambda_{r}}{\lambda_{r}}$ distribution as a function of redshift $z$.}
    \label{fig:CompRichHisto}
  \end{figure*}


\begin{figure*}
\epsscale{1}
    \plotone{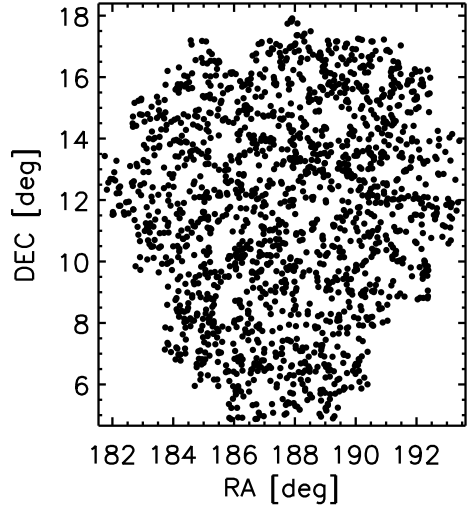} 
    \caption{Spatial distribution of the NGVS cluster candidates in the $104\ \rm deg^2$, detected using four optical bands.}
    \label{fig:spatDistrNGVS4b}
  \end{figure*}

\begin{figure*}
\epsscale{1}
    \plotone{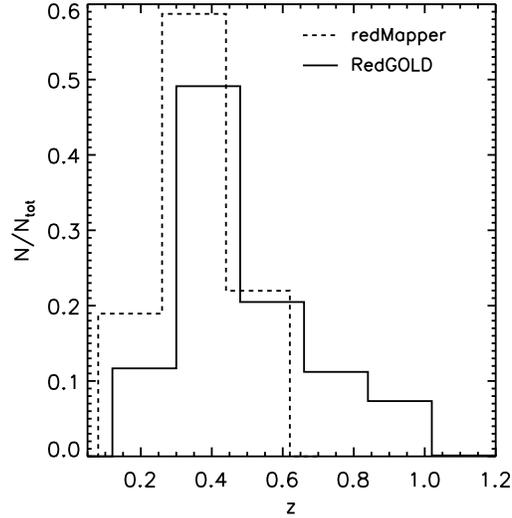} 
    \caption{Redshift distribution of the  {\it RedGOLD$_{wr}$} cluster candidates  (i.e., detected using four optical bands in the entire NGVS field) (solid line). The dashed line represents the redMaPPer detections. Each histogram is normalized to the total number of detections. }
    \label{fig:redsDistrNGVS4b}
  \end{figure*}

\begin{figure*}
\epsscale{1}
    \plotone{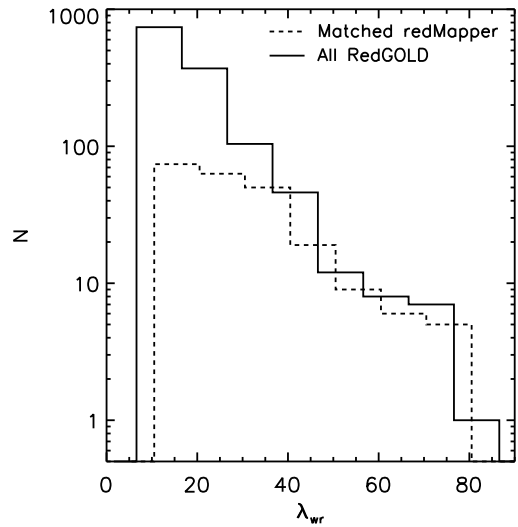} 
    \caption{Richness distribution of the  {\it RedGOLD$_{wr}$} cluster candidates  (solid line) and of the  {\it RedGOLD$_{wr}$} candidates with a counterpart in the redMaPPer catalog (dashed line). Our catalog includes detections with lower richness, i.e. it reaches a lower mass limit with respect to the redMaPPer catalog.}
    \label{fig:CompDistrRichMeRedMNoSDSS}
  \end{figure*}

\begin{figure*}
    \begin{center}
 \epsscale{1.5}
    \plotone{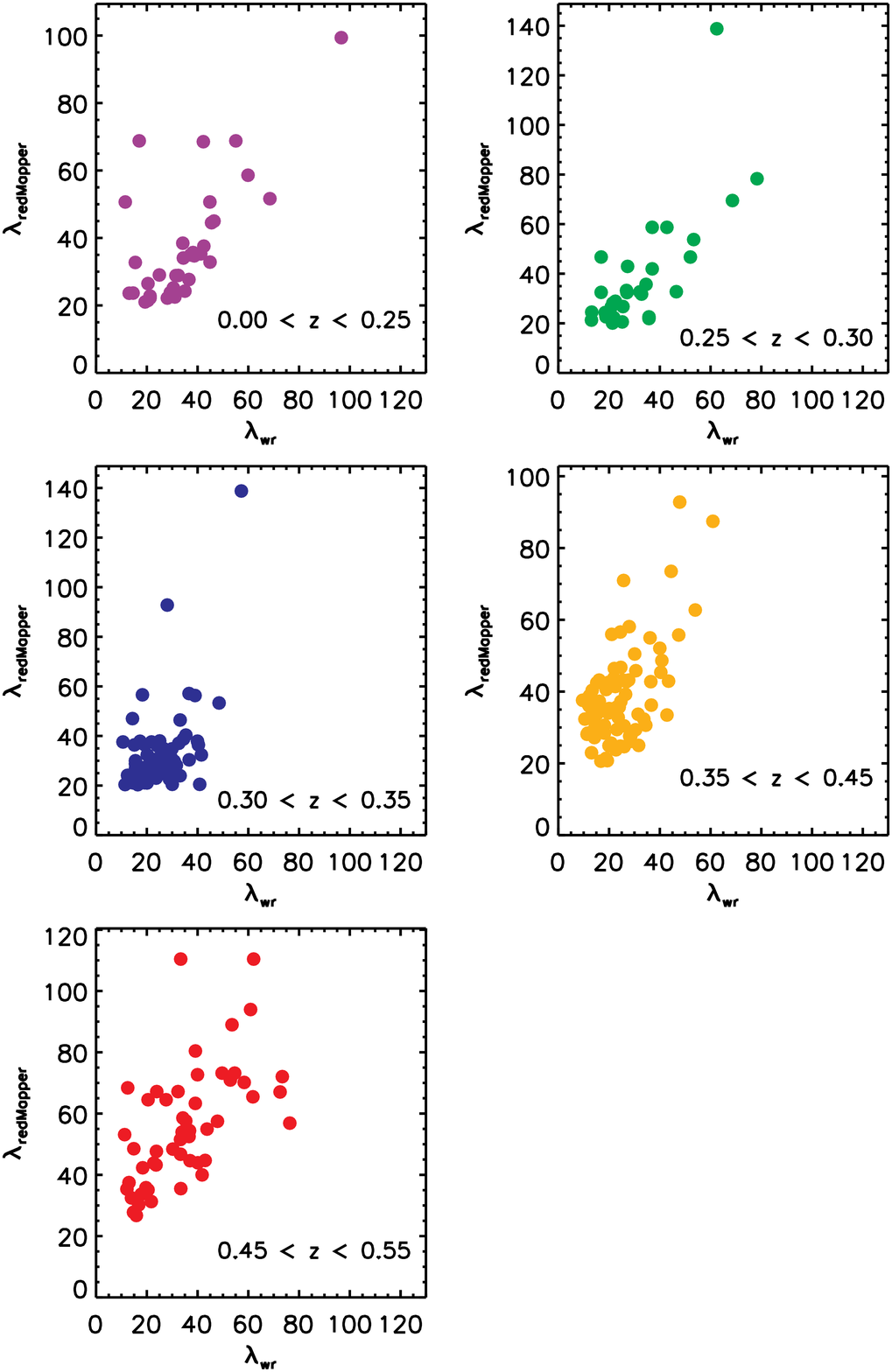}       \end{center}
    \caption{Comparison of the {\it RedGOLD$_{wr}$} richness ($\lambda_{wr}$) vs the redMaPPer richness ($\lambda_{redMaPPer}$)  in different redshift bins as indicated in each panel. }
    \label{fig:CompRich}
  \end{figure*}

\begin{figure*}
    \begin{center}
       \epsscale{1.5}
    \plotone{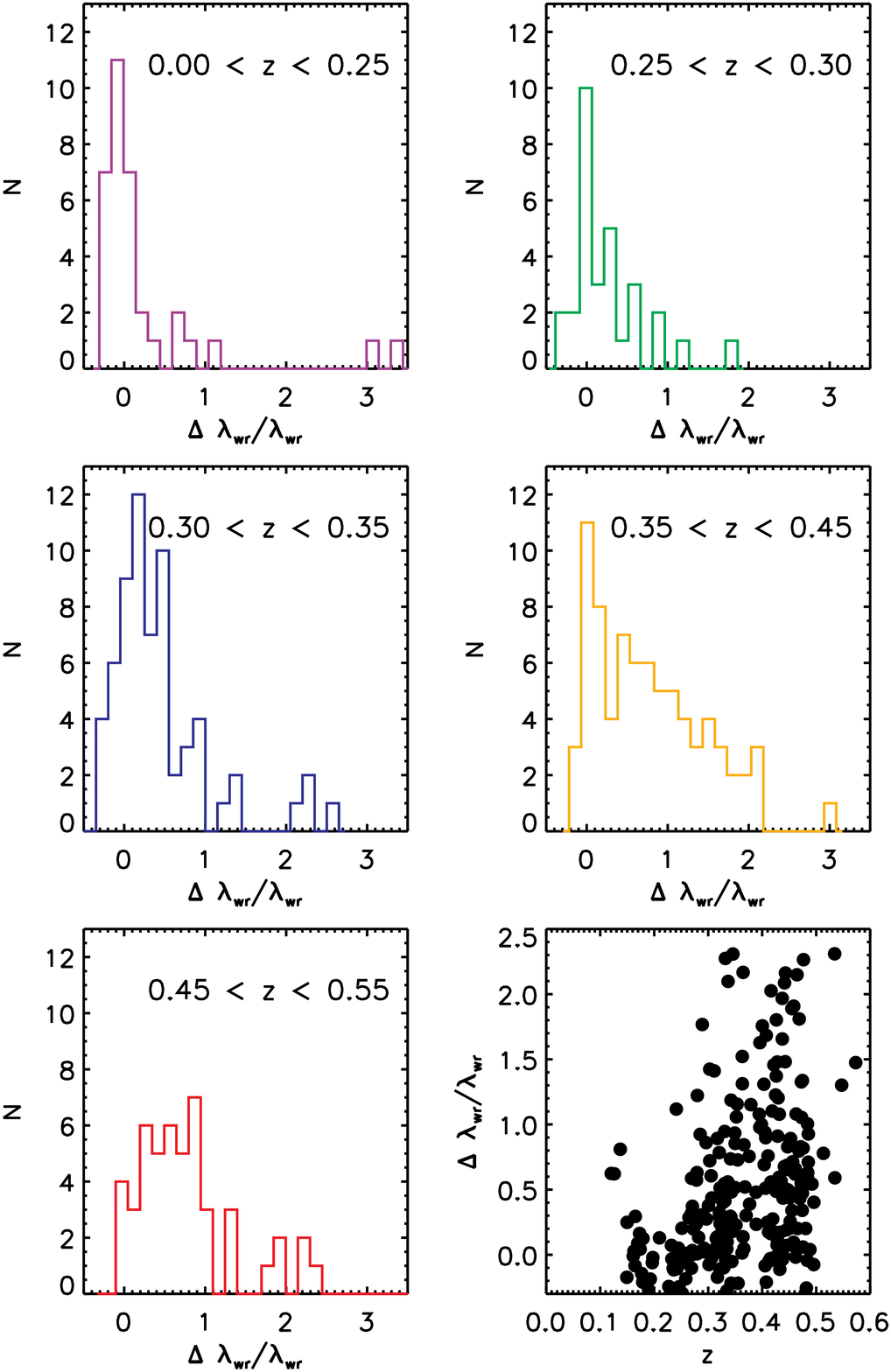}       \end{center}
    \caption{Histogram of  $\frac{\Delta \lambda_{wr}}{\lambda_{wr}} = \frac{\lambda_{redMaPPer}-\lambda_{wr}}{\lambda_{wr}}$, in different redshift bins as indicated in each panel. The bottom right panel shows the $\frac{\Delta \lambda_{wr}}{\lambda_{wr}}$ distribution as a function of redshift $z$.}
    \label{fig:CompRichHisto}
  \end{figure*}

\begin{figure*}
\epsscale{1}
    \plotone{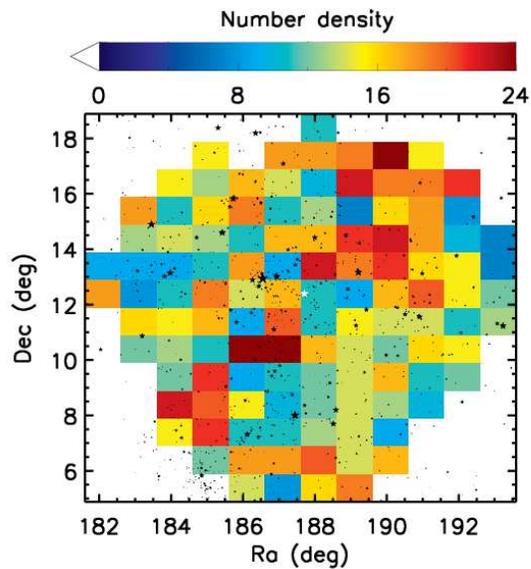} 
    \caption{Density plot of the {\it RedGOLD} detections in each MegaCam pointing.  Different colors correspond to different number of cluster candidates per MegaCam field using four bandpasses, as shown
in the color bar. There is no correlation between the number of detected candidates and the position of the Virgo cluster members. In fact, Virgo bright members mask NGVS areas in which we cannot detect clusters, however this bias is within the variations in the number of detections from pointing to pointing (see text).}
    \label{fig:Dens}
  \end{figure*}

\clearpage

\begin{figure*}
\epsscale{0.8}
    \plotone{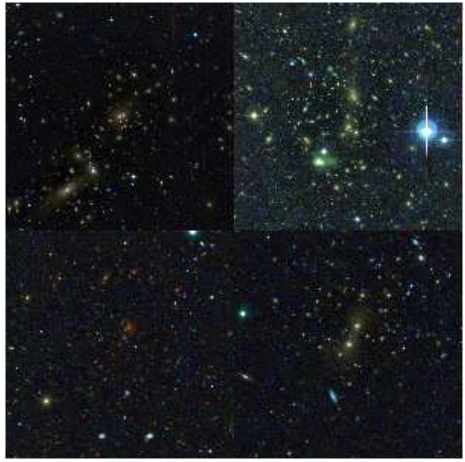} 
    \caption{Optical images of four cluster candidates detected by {\it RedGOLD} in the NGVS area at redshift $z_{cluster} = 0.16$ (left upper panel), $z_{cluster} = 0.27$ (right upper panel), $z_{cluster} = 0.48$ (left lower panel) and $z_{cluster} = 0.25$ (right lower panel). Their detection significance and richness are $\sigma_{det} = 7.7, \lambda=96.6$, $\sigma_{det} = 10.3, \lambda=78.3$, $\sigma_{det} = 19.4, \lambda=76.3$ and $\sigma_{det} = 10.4, \lambda=68.5$.}
    \label{fig:cutout}
  \end{figure*}

\clearpage

\clearpage


\clearpage

\begin{table*}
\begin{center}
\caption{NGVS exposure times, magnitude limits and average seeings in the five optical bands. As the $r$-band data show a large field-to-field depth variation, we provide the magnitude limits range spanned in the different fields \protect\citep[from  Table 1,][]{Raichoor2014}. $mag_{lim}$ is the 5$\sigma$ detection limit in a 2\arcsec aperture \citep[see the details of the estimation in][]{Raichoor2014}.}
\begin{tabular}{cccc }
\tableline\tableline
Filter & exposure time [s] & $mag_{lim}$ [AB] & seeing [''] \\
\tableline
u* & 6402 & $25.60 \pm 0.16$ & $0.83 \pm 0.07$ \\
g  & 3230 & $25.73 \pm 0.13$ & $0.77 \pm 0.08$ \\
r   &  1374 & $[23.56, 25.51] $ & $0.74 \pm 0.14$ \\
i   &   2055 & $24.41 \pm 0.13$ & $0.52 \pm 0.04$ \\
z  & 4466 & $23.62 \pm 0.16$ & $0.70 \pm 0.08$ \\
\tableline
\end{tabular}
\label{tab:NGVSdepth}
\end{center}
\end{table*}

\begin{table*}
\begin{center}
\caption{Median value of the difference $(\lambda_r-\lambda_{wr})/\lambda_r$ and  its standard deviation.}
\begin{tabular}{c c c }
\tableline\tableline
redshift &  $median(\Delta \lambda/\lambda_r)$ & $\sigma_{\Delta \lambda}$  \\
\tableline
$0.1\le z<0.2$ & 0.06 & 0.24 \\
$0.2\le z<0.3$ & -0.04 & 0.11\\
$0.3\le z<0.4$ & -0.08 & 0.17\\
$0.4\le z<0.5$ & 0.05 & 0.14 \\
$0.5\le z<0.6$ & 0.38 & 0.23 \\
$0.6\le z<0.7$ & -0.18 & 0.07 \\
$0.7\le z<0.8$ & -0.24 & 0.33 \\
$0.8\le z<1.1$ & 0.01 & 0.01 \\
\tableline
\end{tabular}
\label{tab:lambda}
\end{center}
\end{table*}

\begin{table}
\begin{center}
\begin{tabular}{c c}
\hline
 redshift & median($\Delta \lambda_{r}/\lambda_{r}$) \\
\hline
$z\le 0.25$ & 0.4 $\pm$  0.5\\ 
$0.25 < z \le 0.30$ &  0.0 $\pm$ 0.3\\ 
$0.30 < z \le 0.35$ &  0.3$\pm$  0.5\\ 
$0.35 < z \le 0.45$ &  0.9 $\pm$  0.6\\ 
$0.45 < z \le 0.55$ &  0.6 $\pm$ 0.5\\ 
\hline
\end{tabular}
\caption{Median value of $(\lambda_{redMaPPer}-\lambda_{r})/\lambda_{r}$ in different redshift bins, obtained using a 3~$\sigma$ clipping.}
\label{Tab:CompRichr}
\end{center}
\end{table}

\begin{table}
\begin{center}
\begin{tabular}{c c}
\hline
 redshift & median($\Delta \lambda_{wr}/\lambda_{wr}$) \\
\hline
$z\le 0.25$ & 0.0 $\pm$  0.8\\ 
$0.25 < z \le 0.30$ &  0.1 $\pm$ 0.5\\ 
$0.30 < z \le 0.35$ &  0.3$\pm$  0.6\\ 
$0.35 < z \le 0.45$ &  0.7 $\pm$  0.7\\ 
$0.45 < z \le 0.55$ &  0.7 $\pm$ 0.9\\ 
\hline
\end{tabular}
\caption{Median value of $(\lambda_{redMaPPer}-\lambda_{wr})/\lambda_{wr}$ in different redshift bins, obtained using a 3~$\sigma$ clipping.}
\label{Tab:CompRich}
\end{center}
\end{table}

\clearpage




\end{document}